\documentstyle[twoside,fleqn,espcrc2]{article}

\newcommand{\AmS}{{\protect\the\textfont2
  A\kern-.1667em\lower.5ex\hbox{M}\kern-.125emS}}

\title{Measuring the gluon spin in protons through $\eta'$ central production}

\author{J.-M. Fr\`ere \address{Physique Th\'eorique, CP 225 \\ 
        Universit\'e Libre de Bruxelles, Blvd du Triomphe, B-1050 Bruxelles}%
        \thanks{Directeur de Recherches, FNRS}
        }
       
\begin{document}

\begin{abstract}
The investigation of a proposed glueball filter in proton-proton collisions
led us to show that 
its action is in fact easily understood in terms of kinematics. While the 
procedure proposed stays of interest in reducing background when looking
for centrally produced resonances of specified spin and parity, we stress that it 
can be put to advantage in giving access to the polarization of gluons in the 
inital protons.\end{abstract}

\maketitle

\section{Introduction}

A "glueball filtering" method has been recently advocated \cite{CLO}
to study central production in $pp$ scattering. 
The proposed method consists in studying the $Q_\perp$ behaviour of 
the production cross section. 
Considering the 2 protons in the process $ p(p_1) p(p_2) 
\to p(p_3) p(p_4) X(k)$, the 
momenta transferred at each proton vertex are erspectively: 
$q_1=p_3-p_1 , q_2=p_4-p_2$ while the variable $Q$ is simply the difference of 
these momenta: $Q=q_1-q_2$ and $Q_\perp$ is its projection in the direction 
transverse to the beam.
It has been advocated that glueballs would be produced even at low values of 
$Q_\perp$, where quark states would be suppressed.
Quite surprinsingly however, even known "glue-rich" states, like $\eta$ or 
$\eta'$ were observed to be 
 suppressed at small values of $Q_\perp$
 \cite{WA102}.
We have thus studied this situation \cite{Escri} and found that rahter than
being a glueball filter, the variable $Q_\perp$ merely distinguishes between
the production kinematics of the various spin and parity states.
In particular, as we will develop below, pseudoscalars of low mass can only 
be produced significantly through the fusion of 2 vectors (either fundamental, 
like gluons or photons, or composite, like the nonet of vector mesons), thus the 
announced behaviour merely results from these simple kinematics and 
the nature of the quark-vector couplings in the proton.
Of course, the suggested cut in $Q_\perp$
 stays a good strategy in reducing background when looking for a specific 
state, for exemple the glueball candidates $f_0$ around 1500 MeV.
{\bf More interestingly even, the argument can be turned back, and central 
production
of $\eta$ or $\eta'$  at low $Q_\perp$
 could be used to measure the spin carried by gluons in the initial protons.}

\section{The production process}
The WA102 and NA12 experiments \cite{WA102,NA12} have examined 
the reaction $pp\rightarrow ppX$ where $X$ is a single resonance 
produced typically in the central region of the collision.

Their experimental
set-up allowed them a complete kinematical study, with redundant measurements
of the protons and $X$ momenta.

We will be more particularly interested in the case where $X$ is a $J^P=0^-$
state, notably $\pi^0$, $\eta$ or $\eta\prime$. 
Neglecting heavier tensor intermediaries,
the production of a pseudoscalar resonance through the fusion of two
intermediaries in parity conserving interactions could arise from
scalar-pseudo\-scalar $(SP)$ or vector-axial $(VA)$ fusion if no factor of 
momenta is allowed, or, vector-pseudoscalar $(VP)$, vector-vector $(VV)$ or 
axial-axial $(AA)$ fusion if the momentum variables can be used 
\cite{CLO2,ARE}. 
Since the first axial resonance is rather heavy,  we restrict 
our discussion to $SP$, $VP$ or $VV$ fusion.

In the case of $SP$ fusion, the only pseudoscalar which could be involved in 
the $\pi^0$, $\eta$ and $\eta\prime$ production is the particle itself, but we
also need a low-lying scalar, possibly the ``sigma'' or a ``pomeron''
state. Moreover, due to the absence of any derivative coupling, the observed
suppression of the production cross section at small $Q_\perp$ cannot occur 
since non trivial helicity transfer is needed (see Ref.~\cite{ARE} for
details). 
Such $(SP)$ fusion is thus obviously disfavoured by experiment.
In the case of $VP$ fusion, the $VPP$ coupling involves one
derivative and should obey Bose and $SU(3)$ symmetry. For instance, a 
$\rho^0\pi^0\pi^0$ coupling is well-known to be forbidden. We conjecture that
the argument can be extended to $U(3)$ symmetry (in particular 
$\rho^0\eta\prime\pi^0$), which removes the discussion of $VP$ fusion from our
analysis.
This leaves $VV$ fusion as the only alternative.

Vector-vector fusion is possible through the vector-vector-pseudoscalar $(VVP)$
coupling
\begin{equation}
\label{VVP}
C_{VVP}=\epsilon_{\mu\nu\alpha\beta} q_1^\mu q_2^\nu
\epsilon_1^\alpha\epsilon_2^\beta\ ,
\end{equation}
where $q_1$ and $q_2$ are the momenta of the exchanged vectors with
polarizations $\epsilon_1$ and $\epsilon_2$ respectively. This coupling is 
well known from the anomalous decay $\pi^0\rightarrow\gamma\gamma$. When 
evaluated in the $X$ rest frame with $k=q_1+q_2$ and $Q=q_1-q_2$, it yields
simply
\begin{equation}
\label{VVPXrf}
C_{VVP}=-\frac{1}{2} m_X \vec{Q}\cdot 
(\vec{\epsilon}_1\times\vec{\epsilon}_2)\ ,
\end{equation}
where clearly the difference $\vec Q$ between $q_1$ and $q_2$ 3-momenta
appears now as a factor and we thus expect a suppression at small $\vec Q$.
But this is insufficient in itself to explain the suppression observed at small
$Q_\perp=|{\vec Q}_\perp|$, where ${\vec Q}_\perp$ is defined as the vector
component of $\vec Q$ transverse to the direction of the initial proton beam.

Here, as seen from (\ref{VVPXrf}), the polarizations of the vectors play an
essential role. In particular, in the $X$ rest of frame, 
${\vec\epsilon}_1\times {\vec\epsilon}_2$ must have components in the $\vec Q$
direction, which implies that both ${\vec\epsilon}_1$ and ${\vec\epsilon}_2$
must have components in the plane perpendicular to $\vec Q$, that is, the 
exchanged vectors must have transverse polarization (helicity $h=\pm 1$). 

The emission of vectors from light fermions is, as well known, restrictive
in helicities. In
the high-energy limit the vector only couples to $\bar f_L\gamma^\mu f_L$ and 
$\bar f_R\gamma^\mu f_R$, that is, the helicity of the fermion cannot change.
In the $X$ rest frame, assumed to lie in the central region of the production,
the colliding fermions cannot (unless they were backscattered, a situation
contrary to the studied kinematical region) emit $h=\pm 1$ vectors in the
forward directions, as this would violate angular momentum conservation.

{\bf We thus reach the conclusion that in the above-mentioned kinematical 
situation,
the production of pseudoscalar mesons by two-vector fusion cannot happen if 
$\vec Q$ is purely longitudinal, but requires 
${\vec Q}_\perp\neq{\vec 0}$} \footnote{There is still a loophole: 
${\vec q}_1$ and ${\vec q}_2$ must have transverse components, but in 
a small area of phase space we could still have
${\vec Q}_\perp=({\vec q}_1-{\vec q}_2)_\perp=\vec 0$. The explicit calculation
 shows this is not significant.}.

It is easy to write down the differential cross section for the central 
production of $X$ and the details are given in \cite{Escri}, and the result 
indeed vanishes when $Q_\perp$ goes to zero.
Although it may seen daring to treat the $p$ as a pointlike particle in the
process considered, this approximation of the $ppV$ ($V$ any vector) coupling
seems phenomenologically more reasonable than the use of a quark parton model
when strictly exclusive processes are considered (where $p$ fragmentation is
not allowed for).

 Under the experimental conditions, the differential cross section thus 
simplifies to (for details see \cite{Escri} and references therein).
\begin{eqnarray}
\label{dcsaprox}
\frac{d\sigma}{dQ_\perp dk_\perp dk_\parallel d\varphi}&\simeq&
\frac{1}{(2\pi)^4}\frac{1}{128 W E p}\nonumber \\[1ex]
&\times& \frac{k_\perp Q_\perp}{|(2p-Q_\parallel)(2E-W)-k_\parallel 
\omega|}\nonumber \\[1ex]
&\times& 16(g_{ppV_1} g_{ppV_2} g_{V_1V_2P})^2 E^2 p^2\nonumber \\[1ex]
&\times&\frac{k_\perp^2 Q_\perp^2\sin^2\varphi}
{(t_1-m_{V_1}^2)^2(t_2-m_{V_2}^2)^2}\ ,
\end{eqnarray}

where the suppression at small
$Q_\perp$ is manifest as it is observed experimentally.

Once the expression for the differential cross section is presented, we may
now enter into conjectures about the nature of the vectors exchanged. 

The most obvous candidates, specially for $t_1, t_2\rightarrow 0$ are the photons 
and gluons. However, the kinematical area explored by, for instance, the 
collaboration WA102, makes it impossible to observe the photonic contribution. We
will return later to the gluon contribution, but for the moment, let us just
stress that it is disfavoured by experiment in the exclusive $p p \to p p X$ 
channel. Indeed, while gluon exchange would forbid the central production of 
single
pions, these dominate in fact the observation. 
We thus have to consider in \cite{Escri} the exchange of the lightest  massive 
vectors 
$\rho$, $\omega$ and $\phi$. 
The couplings of each of the vectors to the proton are known, and their coupling
to $\pi, \eta, \eta '$ were obtained along the lines of ref. \cite{BALL}.
Here the comparison to experiment is made difficult by the need to account 
for phenomenological form factors, in particular at the proton vertices.
The differential cross section correctly reproduces both the size of the 
reactions, and the observed low $Q_\perp$ 
suppression -with an amazing similarity to the observed curves- while the later 
decrease for very large $Q_\perp$ can be fitted by an exponential dependence in 
$t_1,t_2$ .Details of this exponential form factor need to be determined from 
experiment; however if the dominant effect indeed takes place at the proton 
vertex, (see hovever for instance \cite{GER}),it should be universal for all $X$ 
considered, namely, $\pi, \eta$, or $\eta '$ .

\section{Gluon scattering and the spin of gluons inside the proton}
As we have seen above, the gluon-gluon scatternig doesn't play a 
predominant role in the exclusive
central production processes as it would lead to a 
large number of $\eta\prime$ and $\eta$ and no $\pi^0$, which is clearly not
the experimental situation \cite{WA102ref2}. 
Most probably, the selection of isolated protons
in the final state is too restrictive for gluon exchange to take place
significantly.
We would like to advocate an extension of the 
present study to non-exclusive processes $pp\rightarrow\tilde p\tilde pX$,
where $\tilde p$ are jets corresponding to $p$ fragmentation, in order to
observe the {\it QCD} equivalent of the production mechanism 
(gluon-gluon fusion). This will have profound inmplications on the low $Q_\perp$
behaviour of the differential cross section? 

In this case indeed, we must distinguish between gluons emitted from the
fermionic partons (and obeying the helicity constraints discussed at the
beginning of the previous section) and ``constituents'' or ``sea'' gluons.
The latter share part of the proton momentum but their helicity is in
no way constrained. Helicity $h=\pm 1$ gluons can then be met even for 
${\vec Q}_\perp=\vec 0$, and in that case we would expect that the production
distributions in $Q_\perp$ could be considerably affected.

{\bf In this possible extension of the experiments, the $\eta\prime$ and $\eta$ 
now
produced at small $Q_\perp$ are sensitive to the polarization of the 
individual gluons in the proton}. 
Such polarization of the individual gluons is
always present independently of the total polarization of the gluons in the
proton, and is in itself not indicative of the fact that a significant 
proportion of the proton spin could be carried by the gluons. 
If such would be
the case however, and a net polarization of the gluons exists, a similar 
experiment conducted with polarized beams or target would 
lead to a difference in the production rates of $\eta\prime$
and $\eta$ at small $Q_\perp$, and provide a direct measurement of this
polarization.
\section{Conclusions}
In summary, we have shown that the experimental evidence of the
suppression at small $Q_\perp$ of the central pseudoscalar production in
$pp$ scattering can be explained if the production mechanism is through the
fusion of two vectors, and that the corresponding $Q_\perp$ cut is merely 
selecting states on a kinematical basis, rather than being a specific glueball
filter.
This cut still proves very useful in extracting particular resonnances from 
the background (in particular the $f_0$). Furthermore, this kinematical study
has put us on the track of an extension, which {\bf would probe directly the 
gluon contribution 
to the proton spin}.

The experimental extension we suggest goes in two steps.
First, we must include the partially inclusive reactions $pp \to jet \ jet \ X$ ,
and check the $Q_\perp$ distribution in this case; contribution from gluon gluon 
fusion into $X$ should signal itself both by favouring the $\eta, \eta '$ over 
the $\pi$, and by populating the low $Q_\perp$ region in proportion to the amount
of $h=\pm 1$ gluons in the proton.
After this, considering polarised beam and target should allow to study the
behaviour of the low $Q_\perp$ contribution as a function of the polarisation, 
and decide thus on the net spin carried by gluons in the  proton.

\section{Acknowledgements}
I wish to thank both my direct collaborators (Pietro Castoldi, Rafel Escribano),
Frank Close
and my experimental colleagues (Freddy Binon, Andy Kirk, Jean-Pierre Stroot, 
Sacha Singovski), but, in the present occasion, more specially Stefan Narison for 
numerous discussions on glueballs or pseudoscalars, and his kind invitation
and hospitality at QCD98.

\end{document}